\documentclass[
    ,final            
  ]
  {aipproc}

\layoutstyle{6x9}

\begin{document}

\title{Photometric/spectroscopic analyses and magnetic activity in young late-type 
stars}


\author{K. Biazzo}{
  address={INAF -- Catania Astrophysical Observatory, via S. Sofia 78, I-95123, Catania, Italy},
  email={katia.biazzo@oact.inaf.it}
}

\author{A. Frasca}{
  address={INAF -- Catania Astrophysical Observatory, via S. Sofia 78, I--95123, Catania, Italy}
}

\author{E. Marilli}{
  address={INAF -- Catania Astrophysical Observatory, via S. Sofia 78, I--95123, Catania, Italy}
}

\author{E. Covino}{
  address={INAF - Capodimonte Astronomical Observatory, via Moiariello 16, I--80131 Napoli, Italy}
}

\author{J. M. Alcal\`a}{
  address={INAF - Capodimonte Astronomical Observatory, via Moiariello 16, I--80131 Napoli, Italy}
}

\author{${\rm \ddot{O}}$. \c{C}akirli}{
  address={Ege University, Science Faculty, Astronomy and Space Sciences Department, T--35100 
  Bornova, Izmir, Turkey}
}

\begin{abstract}
We present preliminary results of a study in progress based on photometric and 
spectroscopic observations of young weak-line T Tauri and post-T Tauri stars just 
arriving on the Zero Age Main Sequence. 
The study is part of a project based on high-resolution spectra 
obtained with FOCES@CAHA (Spain) and SARG@TNG (Spain) and contemporaneous photometry 
performed at Catania (Italy) and TUBITAK (Turkey) observatories.
The main aim is to investigate the topology of magnetic active regions 
at photospheric and chromospheric levels in young single stars. 	
Since our targets are slow rotators, with rotation periods longer 
than about 2 days, we are able to apply the spectroscopic 
technique based on line-depth ratio for the measure of the photospheric temperature 
modulation. These stars, possible members of Stellar Kinematic Groups, display 
emission cores in the Ca{\sc ii} H\&K and IRT lines, as well as a conspicuous 
filling-in of the H$\alpha$ core. Moreover, we derive lithium abundance, 
and measure rotational and radial velocities. In several cases we detect 
a clear rotational modulation of brightness. In this work we briefly mention 
some preliminary results obtained for SAO~51891. We have also developed a 
spot/plage model to be applied to the data deriving the spot parameters 
(filling factor and temperature) and recovering information about the 
chromospheric inhomogeneities (flux contrast and filling factor).  	
This study will contribute to explore the correlations between global 
stellar parameters and spot/plage characteristics in stars with different 
activity level and evolutionary stage.

\end{abstract}

\maketitle


\section{Introduction}
Stars arriving on the Zero-Age Main Sequence (ZAMS) are in an important evolutionary phase because 
they start to spin up getting free from their circumstellar disks which can begin to condensate giving 
rise to proto-planetary systems. At the same time, if they rotate fast enough, they can undergo a reduced 
effect of angular momentum lost organizing magnetic activity. In addition, young stars belonging 
to Stellar Kinematic Groups (SKGs) are important for a better understanding of the star formation history 
in the solar neighborhood. 
For these reasons, we consider mandatory to investigate the surface topology of magnetic active regions 
(e.g., photospheric spots and chromospheric plages) on young stars possible members of SKGs. Thus, 
we selected stars with high lithium content and low rotational velocity ($<25$ km/s) to be able to 
apply as photospheric diagnostics, besides the light-curve, the line-depth ratios (LDRs). 
The use of contemporaneous light and temperature curves (from LDRs) allows us to derive spot 
temperature and size ($T_{\rm sp}$, $A_{\rm rel}$; Frasca et al. 2005). As chromospheric 
diagnostics, we analyzed the variability of H$\alpha$, helium, and calcium lines as already done for 
intermediate- and high-level active stars (Biazzo et al. 2007b; Frasca et al. 2008). 

In this paper we restrict our attention to the description of the stellar sample and the general outline 
of our project. Moreover, we present some preliminary results for SAO~51891.

\section{The sample}
In 2006-2007 we conducted a photometric and spectroscopic campaign to observe young late-type stars 
in the PMS phase or just approaching the ZAMS. Our aim was to investigate the topology of their magnetic 
active regions at photospheric and chromospheric levels. 
In Table~\ref{tab:star_sample} we list the stellar sample and summarize the journal of spectroscopic and 
photometric observations conducted, respectively, with the high-resolution spectrographs 
FOCES@CAHA\footnote{Calar Alto Observatory (Spain)} and SARG@TNG\footnote{Telescopio Nazionale Galileo (Spain)}, 
and with the photometer at OACt\footnote{Catania Astrophysical Observatory (Italy)} and the 
CCD camera at TUG\footnote{TUBITAK National Observatory (Turkey)}.
%
The columns list star name, year, period, number of nights, and observing site. 
For the spectroscopy, we also indicate the number of spectra. Some of our targets (namely SAO~51891, 
QT~And, and HD~285840) are part of the {\it Spitzer} Legacy Program ``Formation and Evolution of Planetary 
Systems'' (FEPS; Meyer et al. 2006), which is a comprehensive study of the evolution of gas and dust in the 
circumstellar environment of stars spanning a wide range of ages, from the pre-main sequence stage to 
the solar age. This makes the study of these stars even more interesting.\\

\footnotesize
\begin{table}[h]
\begin{tabular}{l|crlcr|clrc}
\hline
    \tablehead{1}{l}{b}{Star name}
  & \tablehead{1}{l}{b}{Year}
  & \tablehead{1}{l}{b}{Period}
  & \tablehead{1}{l}{b}{N}
  & \tablehead{1}{l}{b}{Spectrograph}
  & \tablehead{1}{l}{b}{\#}
  & \tablehead{1}{l}{b}{Year}
  & \tablehead{1}{l}{b}{Period}
  & \tablehead{1}{l}{b}{N}
  & \tablehead{1}{l}{b}{Obs} \\
\hline
SAO~51891         & 2006& Aug 13--16 & 4 & FOCES@CAHA &  11 & 2006& Aug 14--21   & 7 & OACt \\
HD~171488         & 2006& Aug 13--16 & 4 & FOCES@CAHA &  11 & 2006& Aug 14--21   & 8 & OACt \\
HD~177596         & 2006& Aug 13--16 & 4 & FOCES@CAHA &  9  & 2006& Aug 14--21   & 8 & OACt \\
SAO~108142        & 2006& Aug 13--16 & 4 & FOCES@CAHA &  11 & 2006& Aug 14--21   & 8 & OACt \\
QT~And            & 2007& Jan 1--3   & 3 & SARG@TNG   &  7  & 2006& Nov 16--18   & 3 & OACt \\
                  &     &            &   &            &     & 2007& Jan 10--15   & 6 & OACt \\
                  &     &            &   &            &     & 2007& Jan 28--31   & 4 & TUG  \\
                  &     &            &   &            &     & 2007& Feb 7--12    & 6 & TUG  \\
HD~21845          & 2007& Jan 1--3   & 3 & SARG@TNG   &  11 & 2006& Nov 16--18   & 3 & OACt \\
                  &     &            &   &            &     & 2007& Jan 10--15   & 6 & OACt \\
                  &     &            &   &            &     & 2007& Jan 28--31   & 4 & TUG  \\
                  &     &            &   &            &     & 2007& Feb 7--12    & 6 & TUG  \\
BD+45 598         & 2007& Jan 1--3   & 3 & SARG@TNG   &  7  & 2006& Nov 16--18   & 3 & OACt \\
                  &     &            &   &            &     & 2007& Jan 10--15   & 6 & OACt \\
                  &     &            &   &            &     & 2007& Jan 28--31   & 4 & TUG  \\
                  &     &            &   &            &     & 2007& Feb 7--12    & 6 & TUG  \\
HD~285840         & 2007& Jan 1--3   & 3 & SARG@TNG   &  11 & 2006& Nov 16--18   & 2 & OACt \\
                  &     &            &   &            &     & 2007& Jan 10--15   & 6 & OACt \\
RX J0517.9        & 2007& Jan 1--3   & 3 & SARG@TNG   &  9  & 2006& Nov 16--18   & 2 & OACt \\
~~~~~~~~$-$0708   &     &            &   &            &     & 2007& Jan 10--15   & 6 & OACt \\
                  &     &            &   &            &     & 2007& Jan 28--31   & 4 & TUG \\
                  &     &            &   &            &     & 2007& Feb 7--12    & 6 & TUG \\
BD+65 601         & 2007& Jan 1--3   & 3 & SARG@TNG   &  12 & 2007& Jan 10--15   & 6 & OACt \\
                  &     &            &   &            &     & 2007& Jan 28--31   & 4 & TUG \\
                  &     &            &   &            &     & 2007& Feb 7--12    & 6 & TUG \\
HD~237475         & 2007& Jan 1--3   & 3 & SARG@TNG   &  11 & 2007& Jan 10--15   & 6 & OACt \\
                  &     &            &   &            &     & 2007& Jan 28--31   & 4 & TUG \\
                  &     &            &   &            &     & 2007& Feb 7--12    & 6 & TUG \\
\hline
\end{tabular}
\caption{Star sample.}
\label{tab:star_sample}
\end{table}
\normalsize

\subsection{Magnetic stellar activity}
Using photometric data and applying the spectroscopic method based on LDRs 
(Gray et al. 1991, Catalano et al. 2002, Biazzo et al. 2007a) we were able to detect rotational 
modulation of the luminosity and the effective temperature. This allows us to study the stellar 
activity at photospheric level (Biazzo et al. 2008).

\begin{figure}
  \includegraphics[height=.16\textheight]{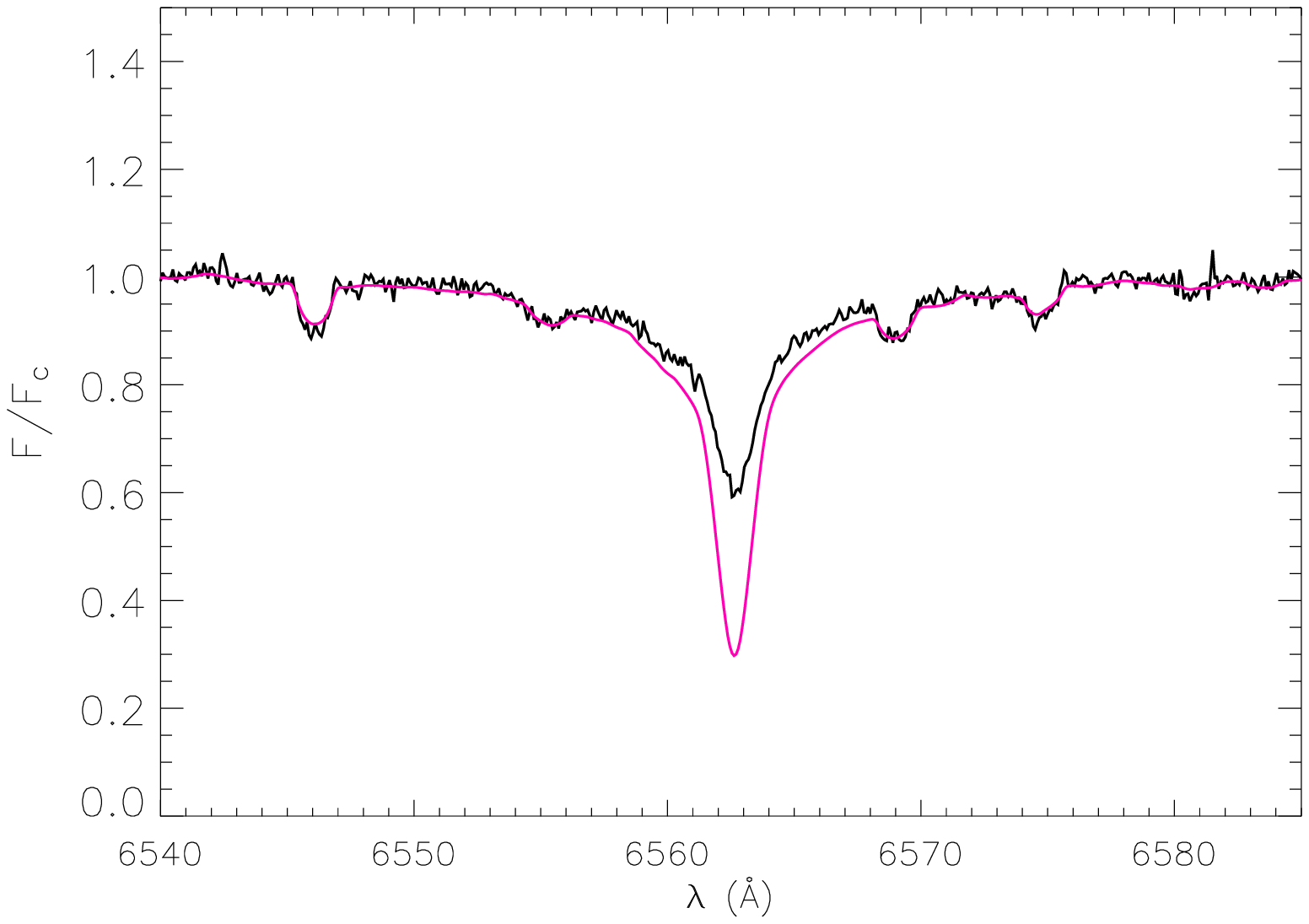}
  \includegraphics[height=.16\textheight]{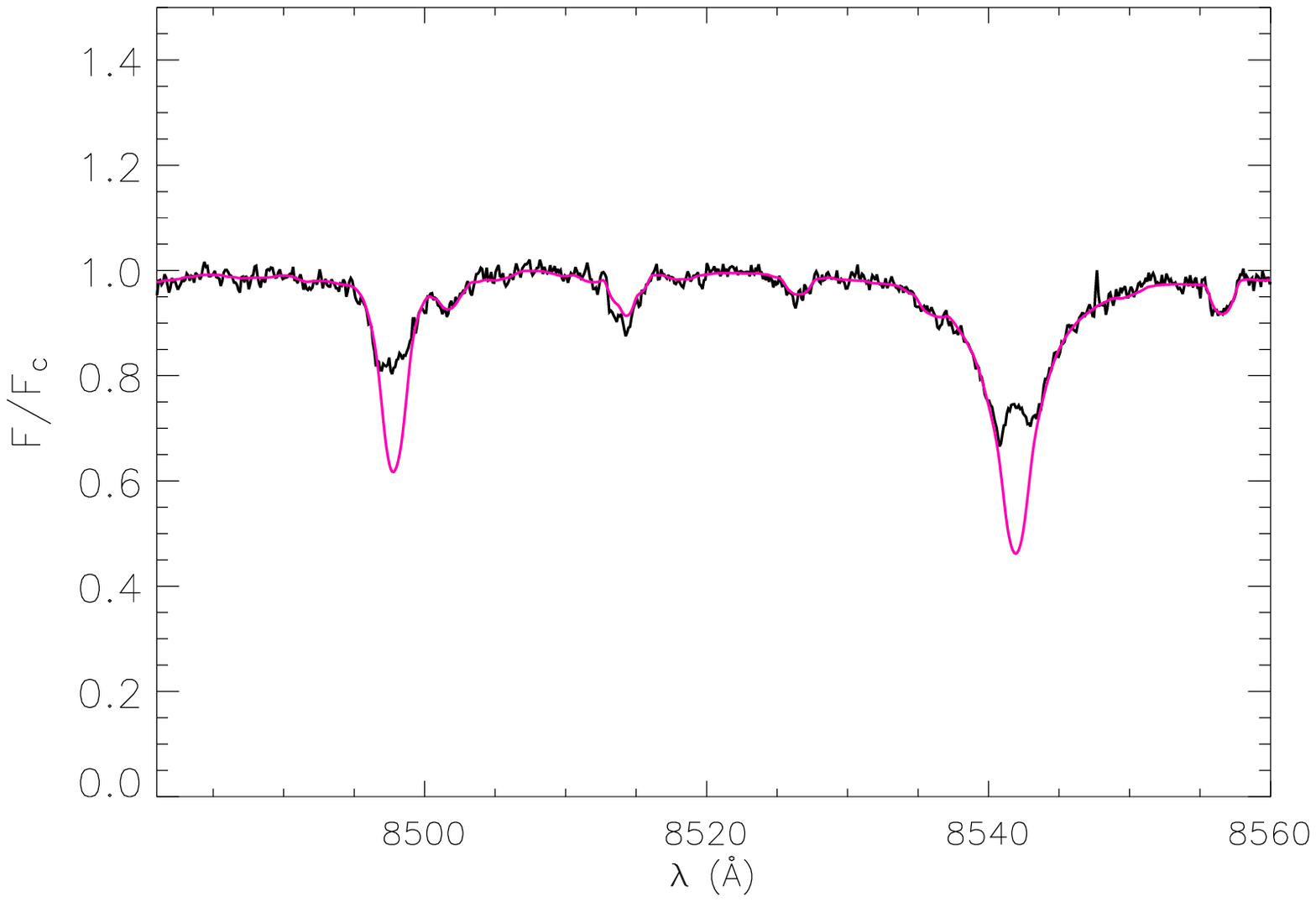}
  \caption{Example of observed (thick line) spectrum of HD~171488 in two spectral regions, together with the 
  non-active template spectrum (thin line).}
\label{fig:spectra_HD171488}
\end{figure}

\begin{figure}
  \includegraphics[height=.16\textheight]{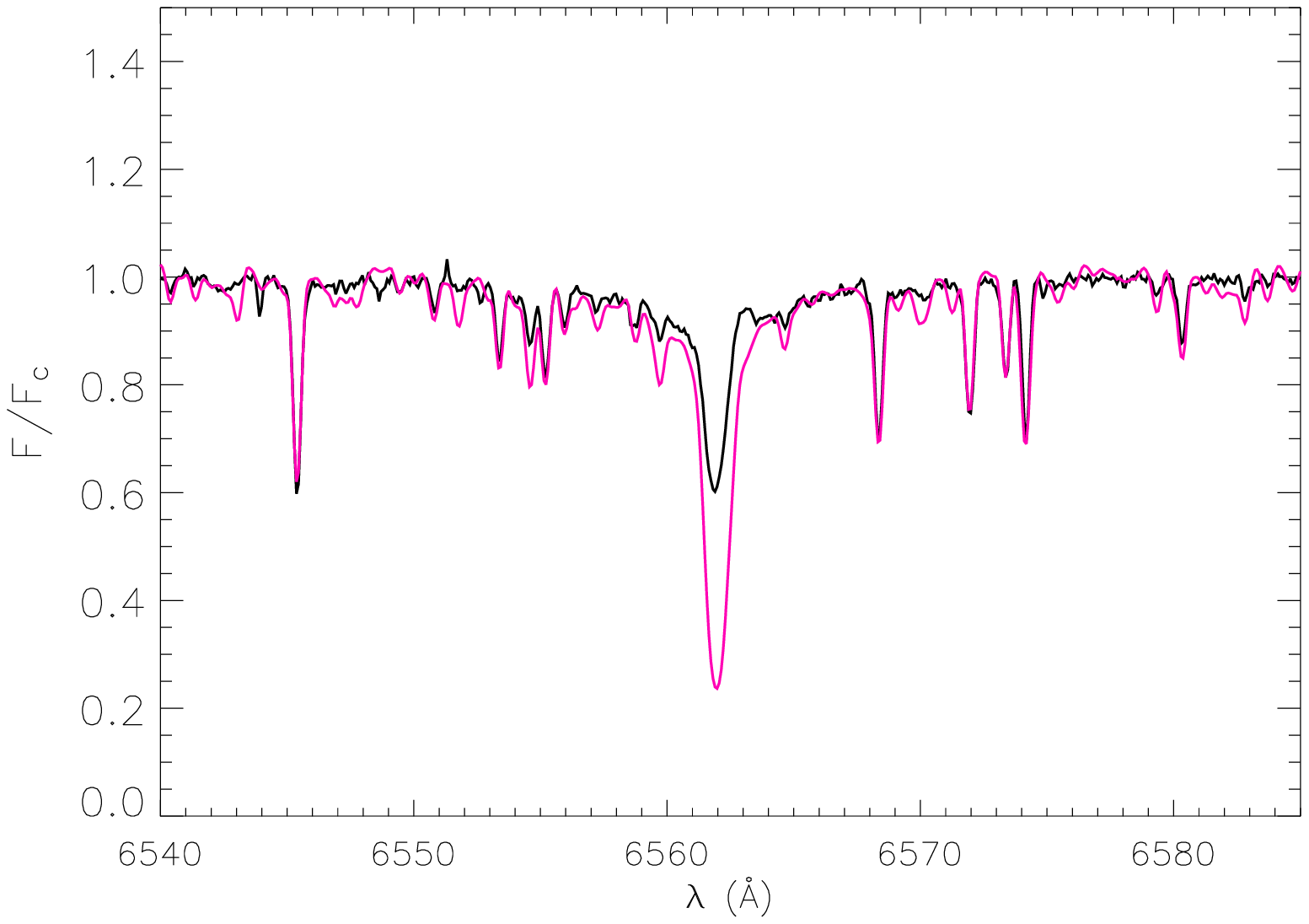}
  \includegraphics[height=.16\textheight]{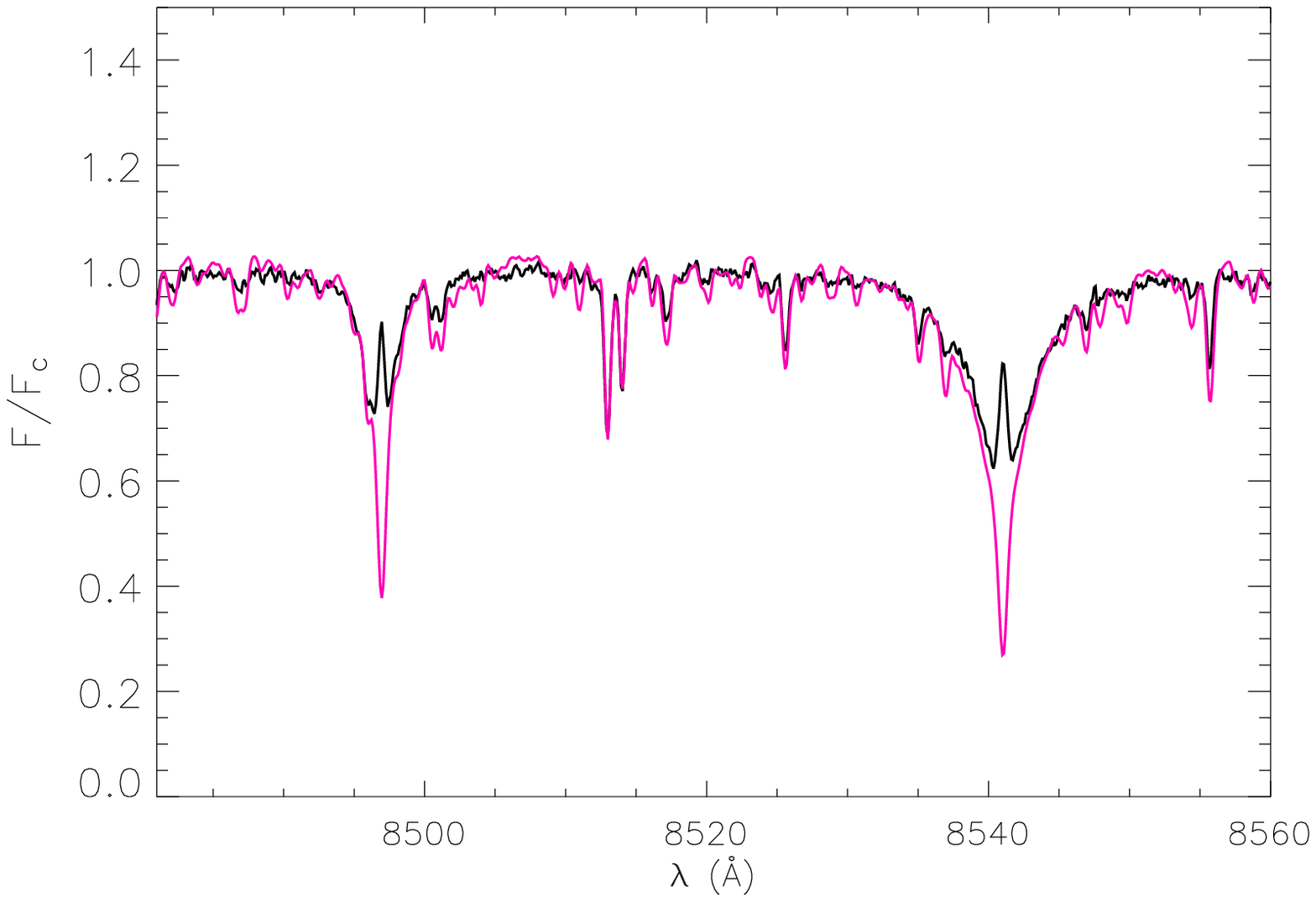}
  \caption{Example of observed (thick line) spectrum of SAO~108142 in two spectral regions, together with the 
  non-active template spectrum (thin line).}
\label{fig:spectra_SAO108142}
\end{figure}

The He{\sc i}-D3, H$\alpha$, Ca{\sc ii} H\&K, H$\epsilon$, and Ca{\sc ii} IRT lines, formed at different 
atmospheric levels, allow us to analyze the chromospheric emission by means of the spectral subtraction 
technique (e.g., Frasca et al. 1994). Figures~\ref{fig:spectra_HD171488}, \ref{fig:spectra_SAO108142} 
show portions of spectrum of HD~171488 and SAO~108142 in the H$\alpha$, and Ca{\sc ii} IRT spectral regions, 
with the non-active template superimposed. The H$\alpha$ and the Ca{\sc ii} IRT profiles are filled-in by 
emission, with the latter displaying a central reversal (very clear in SAO~108142) and suggesting a 
contribution to the total chromospheric losses (Bus\`a et al. 2007).

\section{Conclusions and future work}
From our photometric and spectroscopic observations we are going to study the distribution of inhomogeneities 
at photospheric and chromospheric levels in a sample of young stars. Then, we 
are also developing a spot/plage model to reproduce the 
data deriving the main spot parameters ($T_{\rm sp}$, $A_{\rm rel}$) and obtaining 
information about the chromospheric plages (flux contrast and filling factor). 
Our spot/plage model applied to SAO~51891 (Biazzo et al. 2008) points towards bigger and warmer spots 
($T_{\rm sp}$ closer to the photospheric temperature value) compared to wildly-active main-sequence 
stars (Biazzo et al. 2007b). We are currently refining our spot/plage model and planning to 
observe our targets in other epochs in order to draw firmer conclusions.

This study will contribute to explore the relationships between global stellar parameters 
($T_{\rm eff}$, $\log g$, [Fe/H], age) and spot/plage properties in stars with different activity 
level and evolutionary stage.

\begin{theacknowledgments}
We thank the OACt, TNG, CAHA, and TUG teams for the assistance during the observations. 
This work has been supported by the Italian {\em Ministero dell'Istruzione, Universit\`a e Ricerca} (MIUR) 
which is gratefully acknowledged. This research has made use of SIMBAD and VIZIER databases, operated at CDS, 
Strasbourg (France).
\end{theacknowledgments}



\bibliographystyle{aipproc}   




\end{document}